# Security Design Patterns in Distributed Microservice Architecture

Chaitanya K. Rudrabhatla

*Executive Director – Solutions Architect*
*Media and Entertainment domain*
*Los Angeles, USA*

*Abstract*- **Micro service architecture has revolutionized the landscape for the development of web and mobile applications alike. Due to the stateless nature and loose coupling involved in the design of micro services, native mobile applications can be developed by utilizing the same backend services which feed the inputs to the web application front ends. Extending the same concept, a plethora of automated devices, thanks to the advancements in the field of IOT, have come into existence which can feed on the same set of micro services. This concept of build once and utilize for many use cases has become a new norm in the enterprise design patterns. To handle the horizontal scalability needs of so many calling clients, significant advancements have been made on the containerization and their orchestration strategies on the public cloud platforms. However, scalable design techniques have led to the increased exposure of backend services to unwanted entities. This broadened the attack surface and also the risk. On top of it the mix of heterogeneous technologies in MSA, their distinct logging strategies, makes the central logging difficult, which in turn loosens the security. Additionally, the complexity around building the resilience for fault tolerance across the decentralized networks, adds to the security loop holes. The simple security designs which were once used with traditional web applications cannot be used for Microservice based applications. This paper articulates the innovative approaches of handling the security needs involved in protection of distributed services in Microservice architecture.**

I. INTRODUCTION

Traditionally all the web applications were designed as monolithic in nature. In that traditional setup, MVC (Model, View Controller) was the most popular paradigm. Application would have a web server where all the static content like images, java script and HTML was hosted. In an internet based web application, these web servers are deployed in the DMZ (demilitarized zone) network layer. The web application code itself is compiled and packed as an enterprise archive file(.ear) or a web archive file (.war) file. This archive file is deployed on the enterprise application servers like IBM web sphere, BEA web logic, JBOSS EAP server or Apache Tomcat server to name a few. These servers are deployed in the organization's internal network and shielded from being directly exposed to internet. This archive file would communicate with the backend databases, other data stores, or would integrate with third party services using web services or asynchronous integration mechanisms. This application deployment pattern can be visualised in the Figure 1 below.

As it can be understood from the diagram that the number of entry points to the backend layer are very limited in a monolithic design structure. Not every layer or a component of the application is exposed to the outside world and none of them would accept requests directly from outside. In a typical web application such as the one in Figure 1, all the input requests are scanned for security at the web server level by using a SiteMinder agent which can connect to LDAP systems. Once the request is authenticated it then passes down to the application level. At this level the requests are further filter by a servlet filter. This security check verifies whether the incoming request is associated with a valid session identifier or not. If not, it challenges the incoming request session to authenticate first. Further authorization checks may be in place to validate that the requesting user has the necessary roles and permissions to access the intended resources. The intercepting servlet filter carries out such tasks centrally to make sure that only authenticated and authorized requests are dispatched to the corresponding resources. Internal layers and code blocks may not be concerned about the legitimacy of the requests; they can be confident that if a request ends up there, all the security checks would have already been taken care of. On top of it, since these are stateful applications, the internal components can acquire the user information at any point of time by querying the web session in hand. It is the servlet filer which injects the incoming user information into the session variables during the initial filter process. However, the security design cannot be so simple in case of distributed Microservice based architecture. With the advent of cloud platforms and discovery of containerization techniques using docker etc. and orchestrating software like Kubernetes or docker swarm etc. have complicated the problem even further as explained in the section below.

II. DESIGN COMPLEXITY IN SECURING MSA

Micro services architecture deals with breaking down the application to smaller and independent deployable units for getting a number of benefits [1] [2] including but not limited to –

- service isolation
- separation of concerns
- horizontal scalability and containerization
- quicker build cycles
- easier maintenance
- Isolation of failures
- cloud portability etc.

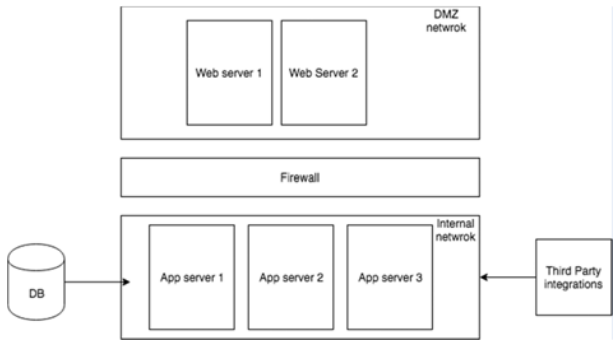

Figure 1 - Traditional Monolithic MVC application design

This design has helped to solve many issues which a traditional monolithic application could not have aimed for. But at the same time, it introduced the following security challenges –
(1) since the logic layer and middle tier of the application is split into so many smaller and independent pieces, the attack surface has significantly increased. (2) Also, due to the lose coupling of the front end and backend layers, it permitted multiple front-end systems to talk to the backend services. This has also broadened the attack surface. (3) with the rise of Internet of Things (IOT) many automated devices started interacting with the micro service layers. Since there is no human intervention involved in these data transfers, authentication and authorization are not straight forward.
Due to these reasons, legacy techniques which were once used for protecting the monolithic web applications can no longer be used in case of micro service architectures. Since the back end micro services can be widely distributed, they can be present in a single datacentre or a cloud environment or a combination of data canters and public clouds [3]. On top of it, unlike monolith the front end and backend are totally detached, and the backend services can now be written in heterogeneous technologies [4]. For example- few services might be running on Java language and other might be on Node.JS or Python. The microservice architecture at a high level can be visualized from Figure 2 given below.

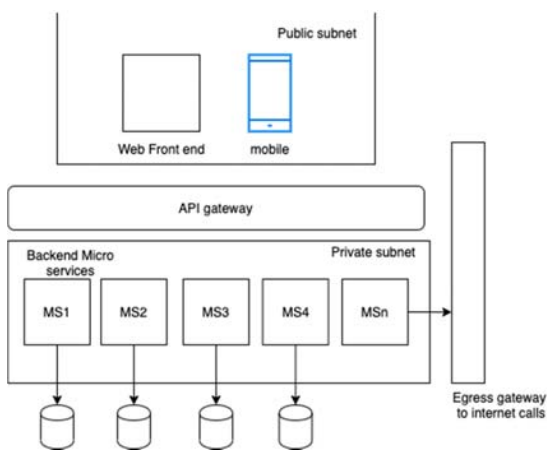

Figure 2 – Microservice architecture design

### III. SECURING THE MICRO SERVICES

Traditionally security teams have always focussed on protecting the databases, network and infrastructure layers. But with the introduction of distributed MSA architectures[8][9], the attack vector has shifted to service side of applications. This is evident from some of the major cyber-attacks which occurred in the recent times. A robust security design for MSA needs to focus on the following main areas –

#### A. Securing the attack surface

*Problem* – As explained in the earlier sections the attack surface has snowballed in the recent times due to the proliferation of the services using their containerization designs and auto-scaling features on public clouds [10]. Front end systems are isolated and can most likely be more than one, unlike the traditional design where backend and front end are packed as one unit. Old security designs using web access management agents like SiteMinder running on top of webservers cannot be applied to distributed microservices with segregated from the front-end systems [5][6].

*Solution* - For securing the distributed microservices which are segregated from the front-end systems, there needs to be an authentication and authorization mechanism which can generate the tokens which are short lived. The movement towards this approach, caused software teams to look for security tokens that worked with RESTful and JSON formats. Some of the token technology that evolved from this requirement include JSON web token (JWT) and OAuth 2.0. OpenID and Auth0 are the 2 important protocols which can solve this issue. Also, to reduce the exposure of services to the callers, there needs to be a single point of contact which can route the calls to the appropriate backend service. This gave way to API gateways. Figure 3 describes the high-level design for using the token-based mechanisms using API gateways.
Here is a high-level flow which explains how the token mechanisms can protect the backend services –

- User (Resource owner) requests the protected resource which is hosted on a microservice (resource server)
- Front end (client) checks if the valid token is present. If not, it sends the user to a login page which has multiple authentication challenges depending on the organizational needs.
- Once user enters the authentication credentials, a call is made to the authorization server via the API gateway.
- Authorization server authenticates the user credentials by validating it with the identity provider. This can be any LDAP system present in the organization.
- If the credentials are valid, the authorization server returns a short-lived access token and a refresh token to the front-end system.
- Front end stores the refresh token. It calls the resource server again by passing the access token in the request header.
- API gateway has a filter mechanism, which first validates the access token by firing an authorization check with the authorization server.

- If the check passes, the request is sent to the resource server for delivering the request needed by the user.
- If in case the access token is invalid, the authorization server denies access, which is handled by the front-end system as a message on the login screen.
- In case, the access token is expired, the front-end system can request a new access token from the authorization server by passing the refresh token.

If the refresh token is invalid too, then the front end can request the user to re-login to the application.

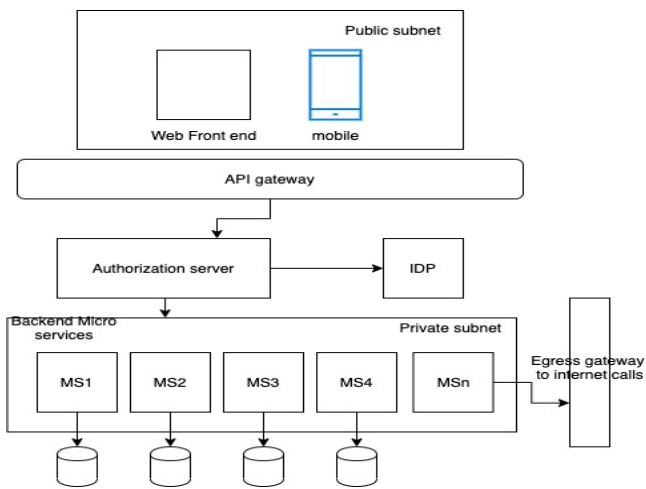

Figure 3 – MSA security using STS and API gateway.

### B. Defence-in-Depth implementation.

*Problem* – Backend services are of various sensitivities and priorities. It cannot be one size fits for all strategy for all services. Depending on the micro service the level of security should be changed. For example, there may be public read APIs which can be exposed to all callers. There can be other services which can be exposed once the caller is authenticated and there can be some other premium services or admin services which not only need to be authorized based on the user profile but also be protected carefully. Simple STS with API gateway cannot be a solution for this need.

*Solution* – For handling this kind of services, a strategy like defense in depth needs to be employed. Defense in depth is a concept in which multiple layers of security controls (defense) are placed throughout the system rather than a single defense system across all layers. For the use case where extra level of security needs to be implemented for premium or admin services, there can be an additional layer of security in the form of private API gateway (not exposed to callers) with an additional filter and authorization scheme. If a caller needs to reach these inner secure layers, they need to access via façade services and pass an extra level of authorization as shown in Figure 4.

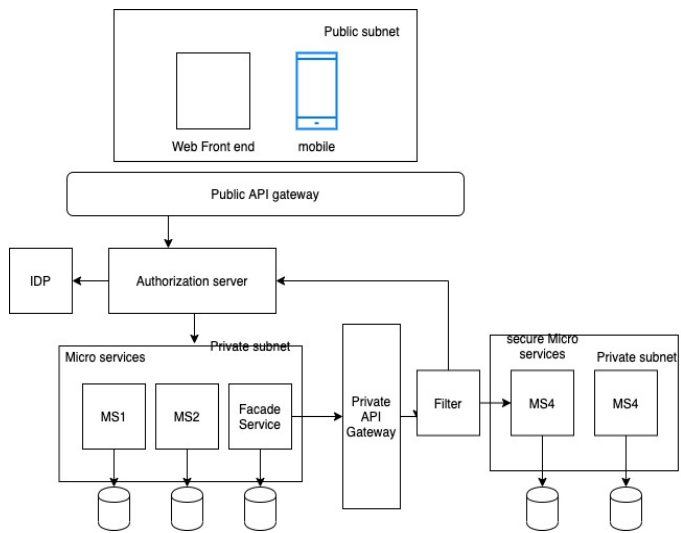

Figure 4 – Defence in depth in MSA security using private API gateway.

### C. Principle of least privilege.

*Problem-* Due to the database per service pattern, each service is independent of the other, including the databases. For a transaction which joins multiple domains, services need to talk to each other either directly or via orchestrating saga patterns. Many a times inter service communication is complex to implement. Such design would need to be intricately done. It is a challenge to narrow down the rules which govern the privileges and permissions for the APIs involved in the services.

*Solution-*APIs need to be designed based on principle of least privilege. Access need to be given only to those users who are not only authenticated but authorized to access the API. API access only need to be granted based on the need. Start with a minimum access and only expand based on the need.

### D. Encryption everywhere.

*Problem-* Services can be decomposed efficiently using various governing principles like Common Closure Principle (CCP) and Single Responsibility Principle (SRP), to achieve the maximum isolation. But the interservice communication is unavoidable as the transactions span across multiple entities. Also, the data is persisted in multiple heterogeneous database technologies based on the needs. Securing the data is a challenge.

*Solution-* Encrypting the data at rest using technologies like Vormetric transparent encryption or by using the secure keys which are rotated once in every few days might solve the security needs for the data at rest. Also, the storage buckets for the files and caching systems should be encrypted using secure passwords. Not only the data at rest but the data in transit should be encrypted using certificates and SSL layers.

*E. Continuous scanning, monitoring, logging -DevSecOps.*

Problem – There are various other problems deeply embedded in MSA. Since all the services are stateless, independent and heterogeneous, there cannot be a uniform logging pattern and framework. This makes threat identification difficult. The other challenge with most development teams is the early identification of the vulnerabilities before they creep to production. And also, designing the security around the containers and various components is a complex challenge.

*Solution-* In case of microservices, it is imperative and essential to design a central log aggregation solution. This would help narrow down the issue faster and avoid corruption to other layers and services. Along with these employing appropriate CAST and DAST scans would mitigate the issues upfront. Also, the methodology of Devops should shift to DevSecOps where Dev, Ops and Security teams make a combined call on micro service architecture rather than dealing with a siloed approach. Constant monitoring and feedback mechanisms may be employed for better results.

## IV. CONCLUSION

This paper outlines some of the important concepts which are needed for securing the distributed service architectures. However, securing the services in a MSA is relatively new concept and would continue to evolve more in the coming years. Lot of research is also being done in the field of AI where the threat can be forecasted based on the oddities in the service requests, responses or interaction patterns. These forecasts can be validated using the machine learning techniques to come up with mechanisms to avoid the infiltration. At the minimum, the service can be shut down by invoking the circuit breakers and alerting the responsible parties. Future scope of work includes devising the innovative techniques to implement the multi factor authentication (MFA) based security in the IOT devices which do not have any human intervention.